\documentclass[12pt]{article}
\usepackage{amssymb} 
\usepackage{amsmath} 

\begin{document}
\font\frak=eufm10 scaled\magstep1
\font\fak=eufm10 scaled\magstep2
\font\fk=eufm10 scaled\magstep3
\font\black=msbm10 scaled\magstep1
\font\bigblack=msbm10 scaled\magstep2
\font\bbigblack=msbm10 scaled\magstep3
\font\scriptfrak=eufm10
\font\tenfrak=eufm10
\font\tenblack=msbm10


\def\biggoth #1{\hbox{{\fak #1}}}
\def\bbiggoth #1{\hbox{{\fk #1}}}
\def\sp #1{{{\cal #1}}}
\def\goth #1{\hbox{{\frak #1}}}
\def\scriptgoth #1{\hbox{{\scriptfrak #1}}}
\def\smallgoth #1{\hbox{{\tenfrak #1}}}
\def\smallfield #1{\hbox{{\tenblack #1}}}
\def\field #1{\hbox{{\black #1}}}
\def\bigfield #1{\hbox{{\bigblack #1}}}
\def\bbigfield #1{\hbox{{\bbigblack #1}}}
\def\Bbb #1{\hbox{{\black #1}}}
\def\v #1{\vert #1\vert}             
\def\ord#1{\vert #1\vert} 
\def\m #1 #2{(-1)^{{\v #1} {\v #2}}} 
\def\lie #1{{\sp L_{\!#1}}}               
\def\pd#1#2{\frac{\partial#1}{\partial#2}}
\def\pois#1#2{\{#1,#2\}}
\def\set#1{\{\,#1\,\}}             
\def\<#1>{\langle#1\rangle}        
\def\>#1{{\bf #1}}                
\def\f(#1,#2){\frac{#1}{#2}}
\def\cociente #1#2{\frac{#1}{#2}}
\def\braket#1#2{\langle#1\mathbin\vert#2\rangle} 
\def\brakt#1#2{\langle#1\mathbin,#2\rangle}           
\def\dd#1{\frac{\partial}{\partial#1}} 
\def\bra #1{{\langle #1 |}}
\def\ket #1{{| #1 \rangle }}
\def\ddt#1{\frac{d #1}{dt}}
\def\dt2#1{\frac{d^2 #1}{dt^2}}
\def\matriz#1#2{\left( \begin{array}{#1} #2 \end{array}\right) }
\def\Eq#1{{\begin{equation} #1 \end{equation}}}

\def\bw{{\bigwedge}}      
\def\hut{{\scriptstyle \land}}            
\def\dg{{\goth g^*}}                                                                                                            
\def\Cdg{{C^\infty (\goth g^*)}}
\def\poi{\{\:,\}}                           
\def\qw{\hat\omega}                
\def\FL{{\sp F}L}                 
\def\hFL{\widehat{{\sp F}L}}      
\def\XHMw{\goth X_H(M,\omega)} 
\def\XLHMw{\goth X_{LH}(M,\omega)}                  
\def\ea{\varepsilon_a}
\def\ep{\varepsilon}
\def\mitad{\frac{1}{2}}
\def\x{\times}  
\def\cinf{C^\infty} 
\def\forms{\bigwedge}                 
\def\onda{\tilde}
\def\orb{{\sp O}}
\def\exp{{\rm exp}}

\def\a{\alpha}
\def\g{{\gamma }}                  
\def\G{{\Gamma}}	
\def\La{\Lambda}                   
\def\la{\lambda}                   
\def\w{\omega}                     
\def\W{{\Omega}}                   
\def\ltimes{\bowtie} 
\def\coth{\rm coth}
\def\arccoth{\rm arccoth}
\def\arcth{\rm arcth}
\def\sh{\rm sh}
\def\arcsh{\rm arcsh}
\def\ch{\rm ch}             
\def\arcch{\rm arcch}
\def\roc{{\tilde{\cal R}}}                       
\def\cl{{\cal L}}                               
\def\V{{\sp V}}                                 
\def\F{{\sp F}}
\def\cv{{{\goth X}}}                    
\def\LG{\goth g}
\def\LH{\goth h}
\def\X{{{\goth X}}}                     
\def\R{{\hbox{{\field R}}}}             
\def\big R{{\hbox{{\bigfield R}}}}
\def\bbig R{{\hbox{{\bbigfield R}}}}
\def\C{{\hbox{{\field C}}}}         
\def\Z{{\hbox{{\field Z}}}}             
\def\N{{\hbox{{\field N}}}}         

\def\ima{\hbox{{\rm Im}}}                               
\def\dim{\hbox{{\rm dim}}}        
\def\End{\hbox{{\rm End}}} 
\def\Tr{\hbox{{\rm Tr}}} 
\def\tr{{\hbox{\rm\small{Tr}}}}                
\def\lin{{\hbox{Lin}}}
\def\vol{{\hbox{vol}}}  
\def\Hom{{\hbox{Hom}}}
\def\rank{{\hbox{rank}}}
\def\Ad{{\hbox{Ad}}}
\def\ad{{\hbox{ad}}}
\def\CoAd{{\hbox{CoAd}}}
\def\coad{{\hbox{coad}}}                           
\def\Rea{\hbox{Re}}                     
\def\id{{\hbox{id}}}                    
\def\Id{{\hbox{Id}}}
\def\Int{{\hbox{Int}}}
\def\Ext{{\hbox{Ext}}}
\def\Aut{{\hbox{Aut}}}
\def\Card{{\hbox{Card}}}
\def\SODE{{\small{SODE }}}


\newtheorem{teor}{Teorema}[section]
\newtheorem{cor}{Corolario}[section]
\newtheorem{prop}{Proposicin}[section]
\newtheorem{definicion}{Definicin}[section]
\newtheorem{lema}{Lema}[section]

\newtheorem{theorem}{Theorem}
\newtheorem{corollary}{Corollary}
\newtheorem{proposition}{Proposition}
\newtheorem{definition}{Definition}
\newtheorem{lemma}{Lemma}

\def\Eq#1{{\begin{equation} #1 \end{equation}}}
\def\R{\Bbb R}
\def\C{\Bbb C}
\def\Z{\Bbb Z}
\def\d{\partial}

\def\la#1{\lambda_{#1}}
\def\teet#1#2{\theta [\eta _{#1}] (#2)}
\def\tede#1{\theta [\delta](#1)}
\def\N{{\frak N}}
\def\Wei{\wp}
\def\Hil{{\cal H}}

\font\frak=eufm10 scaled\magstep1

\def\bra#1{\langle#1|}
\def\ket#1{|#1\rangle}
\def\goth #1{\hbox{{\frak #1}}}
\def\<#1>{\langle#1\rangle}
\def\cotg{\mathop{\rm cotg}\nolimits}
\def\cotanh{\mathop{\rm cotanh}\nolimits}
\def\arctanh{\mathop{\rm arctanh}\nolimits}
\def\wt{\widetilde}
\def\const{\hbox{const}}
\def\grad{\mathop{\rm grad}\nolimits}
\def\Div{\mathop{\rm div}\nolimits}
\def\braket#1#2{\langle#1|#2\rangle}
\def\Erf{\mathop{\rm Erf}\nolimits}

\centerline{\Large \bf A geometric approach to time evolution}

\bigskip

\centerline{\Large \bf operators of Lie quantum systems} 

\vskip 2cm

\centerline{ Jos\'e F. Cari\~nena$^{\dagger}$, Javier de Lucas$^{\dagger}$
  and Arturo Ramos$^{\ddagger}$}
\vskip 0.5cm

\centerline{$^{\dagger}$Departamento de  F\'{\i}sica Te\'orica, Universidad de Zaragoza,}
\medskip
\centerline{50009 Zaragoza, Spain.}
\medskip
\centerline{$^{\ddagger}$Departamento de An\'alisis Econ\'omico, Universidad de
Zaragoza,}
\medskip  
\centerline{50005 Zaragoza, Spain.}
\medskip
\centerline{\today{}}
\vskip 1cm

\begin{abstract}   
Lie systems in Quantum Mechanics are studied from a geometric point of view. In
particular, we develop
 methods to obtain time evolution operators of time-dependent Schr\"odinger
 equations of Lie type
and we  show how these methods explain certain {\it ad hoc} methods used in
previous 
papers in order to obtain exact solutions. Finally, several instances of
time-dependent quadratic Hamiltonian are solved. 
\end{abstract}

\section{Introduction} 
The use of tools of modern differential geometry has been shown to be very
useful in many different problems in physics and in particular Lie groups
and Lie algebras 
have played a prominent r\^ole in the development of Quantum Mechanics.
The main concern of Lie was the integration of systems of  differential equations  admitting
infinitesimal symmetries but as a byproduct of his work 
we have available   a lot of relevant tools to deal with many different
problems not only in differential geometry and 
classical mechanics but also in Quantum Mechanics. 

Our aim here is to show
the efficiency in solving quantum problems of the theory developed by Lie for dealing with systems of
differential equations admitting  nonlinear
superposition rules for solutions and that therefore 
maybe considered as a generalization of non-autonomous  linear
systems. More specifically, we will be mainly interested in finding 
the time evolution operator for quantum systems described by  time-dependent
quantum Hamiltonians which turn out to be the quantum counterpart  of 
 the above mentioned Lie systems of differential equations.

Time-dependent Schr\"odinger equations of Lie type are Schr\"odinger equations of motion for
which the solutions of the equations can be obtained from the
solutions of an equation in a Lie group $G$ related to the Hamiltonian in a
way to be explained in the paper. 
A particular example is the 
harmonic oscillator,  almost ubiquitous in physics, when the mass and angular
frequency are not constant 
 but become real time-dependent parameters \cite{Le67}-\cite{LLS},
may be as a consequence  of
 interaction with 
the environment \cite{{maa4}}. Methods for solving such systems have been developed 
since long time ago, the generalized invariant method  proposed by Lewis and Riesenfeld \cite{LR69}
being a typical and powerful method. Generalizations of such time-dependent harmonic oscillators 
have also been considered and general quadratic systems have been very often studied from different perspectives \cite{PKM94}--\cite{Ja01a}.

This kind of equations appears very often in physics: quantum
optics \cite{{Ab96},{AK91}}, quantum chemistry \cite{KMT}, Paul trap
\cite{WP90}-\cite{NT00}, quantum dissipation \cite{HW98}, fluid
dynamics \cite{Gi98}, etc. In particular, exact solutions of this type of
equations appear many times in the literature
\cite{maa4}-\cite{CL08}. The methods
 used to obtain the solutions of these problems are numerical or reduce  the problem
of finding out the solution of the time-dependent Schr\"odinger equation to
solve certain differential equations using certain \textit{ad-hoc} hypothesis.
This is, for instance,  the way used in the Lewis-Riesenfeld method \cite{Le67, Gu01} or in the
method of unitary transformations \cite{YKUGP97, So00}. 

Following the techniques of \cite{LS}-\cite{CGM06} we develop in
this paper a geometric interpretation in which the solutions of
this sort of Schr\"odinger equations are obtained  in a natural way,
by  solving a system of differential equations like in other previous papers,
 but
without any additional \textit{ad-hoc} hypothesis. Our presentation is
also  an important
improvement
 from the algorithmic point of view, it offers information about the difficulty
 of solving certain Hamiltonians, allows us to solve different physical
 examples at the same footing and provides us with  a frame to explain different 
methods used in the literature. Finally, our method is an improvement because
it shows that Lie's ideas can be applied in the case of  Sch\"odinger equations.

The paper is organized as follows: Section 2 gives a review of  the
 mathematical theory of Lie systems and summarizes the result of the
 main theorem due to Lie \cite{LS}. Some simple examples of Lie systems are also 
given and special emphasis is made on the main property, the possibility
 of relating them with a particular type of equations on a group. 
Section 3 is devoted to present explicit formulas for a method of
 solving such equations which is a generalization of the method
 proposed by Wei and Norman for linear systems \cite{WN1,{WN2}}. The  motion of a 
classical particle under the action of 
a linear potential is analysed from this perspective. Lie systems 
 in Quantum Mechanics are studied in Section 4 and the reduction method 
in Section 5 where the interaction picture is revisited from a geometric point
of view. Some applications of Lie systems in Quantum Mechanics are studied in
Section 6, and more specifically, time-dependent quadratic Hamiltonians. Other examples
 of Lie systems appearing in the literature are  pointed out 
in Section 7. Finally, the  conclusions and outlook are presented in Section 8. 

\section{Lie systems of differential equations}\label{WN}
In this section we will detail some known results about Lie systems that will be applied along this paper.
First, we recall that time evolution of many physical systems is  described  by non-autonomous 
systems of differential equations
\begin{equation}
\frac{dx^i}{dt} = X^i(x,t)\ , \qquad i=1,\ldots,n\,,
\label{tdynsyst}
\end{equation}
 for instance, Hamilton equations, or Lagrange 
equations when transformed to the first order equations by considering momenta or velocities as new variables. In modern geometric terms,
such a system is substituted by a $t$-dependent vector field
$$X=\sum_{i=1}^n  X^i(x,t)\, \pd{}{x^i}\ ,$$
whose integral curves satisfy (\ref{tdynsyst}).

The theorem of existence and uniqueness of solution 
 for such systems establishes that
 the initial conditions $x(0)=(x^1(0),\ldots,x^n(0))$ determine the future evolution. It is also
 well-known that for the simpler case of a homogeneous linear system there is
 a (linear)
 function $F:{\Bbb R}^{n^2+n}\to {\Bbb R}^n$, given by 
 \begin{equation}
x=F(x_{(1)},\ldots,x_{(n)},k_1,\ldots,k_n)
=k_{1}\, x_{(1)}+\cdots +k_{n}\,x_{(n)}\ ,\label{lsr} 
\end{equation}
in such a way that the 
general solution can be written as a linear combination of $n$ independent 
particular solutions, $x_{(1)}(t), \ldots ,x_{(n)}(t)$,
 \begin{equation}
x(t)=F(x_{(1)}(t),\ldots,x_{(n)}(t),k_1,\ldots,k_n)
=k_{1}\, x_{(1)}(t)+\cdots +k_{n}\,x_{(n)}(t)\ ,\label{lsrt} 
\end{equation}
i.e. $x(t)$ given by (\ref{lsrt}) is a solution for any choice of
$k=(k_1,\ldots,k_n)$  and for each set of initial conditions, 
 the coefficients $k=(k_1,\ldots,k_n)$ can be determined. In a similar way, for
an inhomogeneous linear system, there is an affine superposition function 
$F:{\Bbb R}^{n(n+2)}\to {\Bbb R}^n$ given by 
\begin{eqnarray}
&&x=F(x_{(1)},\ldots,x_{(n+1)},k_1,\ldots,k_n) \nonumber\\
&&\quad\quad\quad=x_{(1)}+ k_1(x_{(2)}-x_{(1)})+\ldots+k_n(x_{(n+1)}-x_{(1)})\ , \label{asr}  
\end{eqnarray}
and the general solution can be written
 as the corresponding  affine function of $(n+1)$ independent particular solutions
\begin{eqnarray}
x(t)&=&F(x_{(1)}(t),\ldots,x_{(n+1)}(t),k_1,\ldots,k_n) \nonumber\\
&=&x_{(1)}(t)+ k_1(x_{(2)}(t)-x_{(1)}(t))+\ldots+k_n(x_{(n+1)}(t)-x_{(1)}(t))\ . \label{asrt}  
\end{eqnarray}
Under a non-linear change of coordinates both systems become  
non-linear ones. However, the fact that the general solution is expressible 
in terms of a set of particular solutions is maintained, 
now the superposition function being  no longer linear or affine, respectively.

The very existence of such examples of systems of differential equations 
admitting a non-linear superposition function suggested to Lie the analysis
and characterization of such systems. He arrived
 in this way to the problem of characterizing the systems 
of differential equations for which a superposition function, 
allowing to express the general solution in terms of $m$ particular solutions, 
does exist. The solution of this problem due to Lie \cite{LS}  asserts
that, 
under very general conditions, 
such systems are those which can be written as  
\begin{equation}\label{AL}
 \frac {dx^i}{dt}=b_1(t)\xi^{1i}(x)+\cdots+b_r(t)\xi^{ri}(x)\,,\qquad i=1,
\ldots,n\,,  
\end{equation}
where  $b_1,\ldots,b_r,$ are  $\,r$ functions  depending only on  $t$ and
  $\xi^{\alpha i}$, $\alpha=1,\ldots, r$,  are functions of 
 $x=(x^1,\ldots,x^n)$,
such that the  $r$ vector fields in   ${\Bbb R}^n$ given by
\begin{equation}
X_{\alpha}\equiv   \sum_{i=1}^n\xi^{\alpha i}(x^1,\ldots,x^n)
\pd {}{x^i}\,,\qquad
 \alpha=1,\ldots,r,
\end{equation}
close on a real finite-dimensional Lie algebra, i.e. the vector fields 
$X_{\alpha}$ are linearly independent  
and  there exist $r^3$
real numbers, $f_{\alpha\beta}\,^\gamma$, such that
\begin{equation}
[X_{\alpha},X_{\beta}]=\sum_{\gamma=1}^rf_{\alpha\beta}\,^\gamma X_{\gamma}\ .
\end{equation}
Even if the theorem was not stated with today level of rigour, the result is
essentially true. For an intuitive geometric proof see \cite{CGM}, and for a more geometric
 approach 
and the uniqueness of such superposition rule see \cite{CarRam03,CGM06}.
 
From the geometric viewpoint, Lie systems are those corresponding to 
a $t$-dependent vector field which is a $t$-dependent linear combination of vector fields
$$X(x,t)=\sum_{\alpha=1}^r b_\alpha(t)\, X_{\alpha}(x)\ ,$$ with
vector fields $X_{\alpha}$ closing on a finite-dimensional real Lie 
algebra. 
Many of the applications in physics and mathematics of such Lie 
systems
have been developed 
by Winternitz and coworkers \cite{And80}--\cite{HavPosWin99}.

We have mentioned in this section that homogeneous linear systems are Lie systems. Let us consider
such a system like
 \begin{equation}
\frac {dx^i}{dt}=\sum_{j=1}^nA^i\,_j(t)\, x^j\ , \quad i=1,\ldots,n\,,
\label{hls}
\end{equation}
for which 
 \begin{equation}
 X= \sum_{i,j=1}^nA^i\,_j(t)\, x^j\,\pd{}{x^i}\ , \label{vfhs}
\end{equation}
which is a
linear combination with time-dependent coefficients, 
 \begin{equation}
X= \sum_{i,j=1}^nA^i\,_j(t)\,X_{ij}\ , \label{livfis}
 \end{equation}
of the $n^2$ vector fields
 \begin{equation}
 X_{ij}=x^j\,\pd{}{x^i}\ , \qquad i,j=1,\ldots,n\,,\label{xij}
\end{equation}
for which 
$$ 
 [X_{ij},X_{kl}]=\left[x^j\pd {}{x^i},x^l\pd {}{x^k}\right]
=\delta^{il}\,x^j
\pd {}{x^k}-\delta^{kj}\,x^l\pd {}{x^i}\ ,
$$
i.e.  
\begin{equation}
[X_{ij},X_{kl}]=\delta^{il}\,X_{kj}-\delta^{kj}\,X_{il}\ ,
\end{equation}
which means that the vector fields $\{X_{\alpha}=X_{ij}, \alpha=(i-1)n+j\}$, with $ i,j=1,\ldots,n$,
appearing in the case of a homogeneous system, 
close on a $n^2$-dimensional real Lie algebra isomorphic 
to the ${\goth{gl}}(n,{\R})$ algebra. Actually they are the fundamental vector
fields corresponding to the natural action of $GL(n,\R)$ on $\R^n$.


Another remarkable example, with many applications in physics, 
is that of Riccati equation, which corresponds to $n=1$ and $m=3$ (see e.g. 
\cite{Win83, CarMarNas, CarRam}).

From the practical viewpoint, the most important property
of Lie systems is that as the vector fields $X_\alpha$ appearing in the $t$-dependent 
linear combination defining the system 
$$X=\sum_{\alpha=1}^r b_\alpha(t)\, X_\alpha\ ,
$$
are assumed to close on  a finite-dimensional
 real Lie algebra $\goth g$, if they are complete vector fields, they generate an effective action 
$\Phi:G \times {\Bbb R}^n\to {\Bbb R}^n$ of a connected Lie group $G$ with Lie algebra $\goth g$, see \cite{LM},
on ${\Bbb R}^n$ and if we determine a curve $g(t)$ in $G$ starting from the neutral element, $g(0)=e$, and such that 
\begin{equation}
R_{g^{-1}(t)*g(t)}\dot g(t) =-\sum_{\alpha=1}^rb_\alpha(t)a_\alpha\equiv a(t)\ ,\label{eqingr}
\end{equation}
where   $\{a_1,\,\dots,\,a_r\}$ is a basis of the Lie 
algebra $\goth g$ closing the same commutation relations as the $X_\alpha$,
then, the solution of (\ref{AL}) with initial condition $x(0)$ is given by
$$
x(t)=\Phi(g(t),x(0))\ .
$$ 
The equation (\ref{eqingr}) is sometimes written with an abuse of notation as 
$$\dot g\,g^{-1}=-\sum_{\alpha=1}^rb_\alpha(t)a_\alpha\equiv a(t)\,.
$$ 

In this way, the problem of finding the general solution of (\ref{AL})
is reduced to that of determining the above mentioned curve in $G$ solution of 
(\ref{eqingr}) and starting from the neutral element \cite{CGM, CarRamGra, CarRamcinc}. In the particular case of linear 
systems the group is (a subgroup of) the general linear group $GL(n,\R)$ and 
the action is linear, and therefore, $\Phi(g(t),\cdot)$ provides the time
 evolution operator.

All the preceding  process can be straightforwardly generalized to deal with
Lie
 systems in a general manifold $M$,  the obtained superposition rule being
generally only local.

The remarkable point is that once a Lie system with Lie group $G$  in a manifold $M$
which is a homogeneous space for $G$  has been solved
 on its group $G$ for a curve $a(t)$ in $\LG$ it is possible to obtain the
 solutions of all  other Lie systems with the same Lie algebra $\LG$ and curve
 $a(t)$ in any other homogeneous space $N$ for $G$. Moreover, if the Lie group involved in the problem is solvable
the problem can be solved by quadratures independently of the curve given in
$\LG$. Otherwise, the problem is to  be solved, when possible,  for each curve in $\LG$
separately, and for some specific examples the solution may be explicitly
shown. Of course, if a problem is solved for a certain Lie group $G$ for all curves
in $\LG$ the problem is also solved for any problem given by a Lie subgroup 
$H$ of $G$. 

\section{The Wei--Norman method\label{Wei_Nor_meth}}

In this section we will describe a method 
  to solve directly the equation (\ref{eqingr})
which is 
a generalization of the one proposed by 
Wei and Norman \cite{WN1,WN2} for finding the time evolution operator 
for a linear systems of type
${dU(t)}/{dt}=H(t)U(t)$, with $U(0)=I$ (see also \cite{CarMarNas}). 
We will only give here the recipe of how to proceed, the proof can be found for instance in \cite{CGM, CarRamGra, CarRamcinc}.

The generalization of Wei--Norman method consists on writing
the solution $g(t)$ of (\ref{eqingr}) 
in terms of its second kind canonical coordinates 
w.r.t. the basis $\{a_1,\,\dots,\,a_r\}$ of the Lie 
algebra $\goth g$, for each value of  $t$, i.e.
\begin{equation}
g(t)=\prod_{\alpha=1}^{r}\exp(-v_\alpha(t)a_\alpha)=\exp(-v_1(t)a_1)\cdots
 \exp(-v_r(t)a_r)\ ,\label{factorizeg}
\end{equation}
and transforming the equation (\ref{eqingr})
into a system of differential equations for the unknown functions 
$v_\alpha(t)$. 
The curve $g(t)$ we are looking for is the one given by
the solution of this last system determined by
the initial conditions $v_\alpha(0)=0$ for all 
$\alpha=1,\,\dots,\,r$. 
The minus signs in the exponentials have been introduced for 
computational convenience.  Now, it can be shown 
that using the expression (\ref{factorizeg}) and 
after some mathematical manipulations, equation   (\ref{eqingr})
becomes
the fundamental expression of the Wei--Norman method \cite{CarRamcinc} 
\begin{equation}
\sum_{\alpha=1}^r \dot v_\alpha \left(\prod_{\beta<\alpha} 
\exp(-v_\beta(t) \ad(a_\beta))\right)a_\alpha
=\sum_{\alpha=1}^r b_\alpha(t) a_\alpha\,,
\label{eq_met_WN}
\end{equation}
with  $v_\alpha(0)=0$, $\alpha=1,\,\dots,\,r$.
The resulting system  of differential equations
for the functions $v_\alpha(t)$ is integrable 
by quadratures if the Lie algebra is solvable 
\cite{WN1,WN2},  for instance, for nilpotent 
Lie algebras. Finally, this system of equations
depends only on the structure constants of the Lie algebra. 


As an interesting example,  from the physical point 
of view,  illustrating the possible applications
 of the theory  we can consider the motion of a classical particle under the action of 
a linear potential. Such example has been studied in \cite{CarRam03} and 
is reproduced here for the sake of completeness. This model, with many 
applications in physics, has been often considered both in classical 
and quantum approaches (see, e.g. \cite{CJV,Fe01}) and it has 
recently been studied by
Guedes \cite{Gu01}; its solution using the theory of Lie systems was given in \cite{CarRamcinc}. The  classical Hamiltonian is
\begin{equation}
H_c=\frac{ p^2}{2m}+f(t)\, x\ . \label{Hc}
 \end{equation}
For instance, when $f(t)=q\, E_0+q\,E\,\cos\omega t $, it describes 
the motion of a particle
of electric charge $q$ and
mass $m$ driven by a monochromatic electric field. 

The classical Hamilton equations of motion are
\begin{eqnarray}{\dot x}&=&\frac pm\,,\cr
 {\dot p}&=&-f(t)\,,\end{eqnarray}
and therefore,
the motion is obtained by two quadratures
\begin{eqnarray}
x(t)&=&x_0+\frac{p_0\, t}m-\frac 1m\int_0^tdt'\,\int_0^{t'}f(t'')\, dt''\ ,\\
p(t)&=&p_0-\int_0^tf(t')\,dt'\,.\label{emls}
\end{eqnarray}

In the geometric approach, the $t$-dependent vector field describing the time evolution  is
$$X=\frac pm\,\pd{}x-f(t)\,\pd{}p\ 
$$
which can be written as a linear combination
$X=\frac 1m\, X_1-f(t)\, X_2$,
with
$$
X_1=p\, \pd{}x\ ,\qquad X_2=\pd{}p\ ,
$$
being two vector fields closing  with 
$X_3=\partial/\partial x$ a three-dimensional real Lie algebra
isomorphic to the Heisenberg algebra, namely,
\begin{equation}
[X_1,X_2]=-X_3\ , \qquad [X_1,X_3]=0\ ,\qquad [X_2,X_3]=0\ .\label{Xla}
\end{equation}

The flow of these vector fields is given, respectively, by
\begin{eqnarray}
&&\phi_1(t,(x_0,p_0))=(x_0+p_0\, t,p_0)\,,      \nonumber\\
&&\phi_2(t,(x_0,p_0))=(x_0,p_0+t)\,,            \nonumber\\
&&\phi_3(t,(x_0,p_0))=(x_0+t,p_0)\,.            \nonumber
\end{eqnarray}

In other words, this corresponds to the action of the Lie group of 
upper triangular $3 \times3$ matrices, the Heisenberg group,  
on $\R^2$\ ,
$$
\matriz{c}{\bar x\\ \bar p\\1}=\matriz{ccc}{1&\alpha_1&\alpha_3\\0&1&\alpha_2\\0&0&1}
\matriz{c}{x\\ p\\1}\ .
$$
 
 Let $\{a_1,\,a_2,\,a_3\}$ be a basis
of the Lie algebra with non-vanishing defining relations $[a_1,a_2]=-a_3$. 
Then, the corresponding equation in the group (\ref{eqingr}) becomes in this case 
$$
\dot g\, g^{-1}=-\frac 1 m a_1+f(t)\, a_2\ .
$$
Now, choosing the factorization
$g= \exp(-u_3\, a_3)\,\exp(-u_2\, a_2)\,\exp(-u_1\, a_1)$ and using 
the Wei--Norman formula (\ref{eq_met_WN})
we will arrive to the system of differential equations
$$
\dot u_1=\frac 1m\ ,\qquad \dot u_2=-f(t)\ ,\qquad
\dot u_3-\dot u_1\, u_2=0\ ,
$$
together with the initial conditions
$$u_1(0)=u_2(0)=u_3(0)=0\ ,
$$
with solution 
$$
u_1=\frac tm\ ,\qquad u_2=-\int_0^tf(t')\, dt'\ ,
\qquad u_3=-\frac 1m \int_0^t dt'\int_0^{t'}f(t'')\, dt''\ .
$$

Therefore the motion will be given by
$$
\matriz{c}{x\\p\\1}= \matriz{ccc}{1&\frac tm&-\frac 1m\int_0^t dt'\int_0^{t'}f(t'')\,
 dt''\\0&1&-\int_0^tf(t')\, dt'\\0&0&1}\matriz{c}{x_0\\p_0\\1}\ ,
$$
which reproduces (\ref{emls}). Inverting the matrix we will obtain two
 constants of the 
motion corresponding to the values of the initial conditions
\begin{eqnarray} 
I_1&=&p(t)+\int_0^tf(t')\, dt'\,,\cr
I_2&=&x(t)-\frac tm\left(p(t)+\int_0^tf(t')\, dt'\right) t
+\frac t m\int_0^tdt'\int_0^{t'}f(t'')\, dt''\,,\nonumber
\end{eqnarray}
the first one being the one given  in \cite {Gu01}. 

\section{Schr\"odinger Lie systems in Quantum Mechanics}\label{SLSQM}

In this section we will review a way to generalise Lie's ideas in order 
to use them in the case of Sch\"odinger equations \cite{CarRam03,CarRam05b}. As a new result we show how to apply our method to obtain solutions for quadratic Hamiltonians. Many particular cases of this Hamiltonian can be found in the literature, but obtained by {\it ad hoc} or approximate methods as it will be explained in detail in a next section. In our method, nevertheless, we found an algorithmic way to find them all.

As far as  Quantum Mechanics is concerned, let us first remark that  
the separable complex Hilbert space of states $\cal H$ can be seen as a 
(infinite-dimensional)
real manifold admitting a global chart \cite{BCG}. The Abelian translation 
group provides us with an 
identification of  the tangent space $T_\phi\cal H$ at any point $\phi\in
\cal H$  with  $\cal H$ itself,  the isomorphism being obtained by
associating with $\psi\in\cal H$ the vector $\dot{\psi}\in
T_\phi\cal H$ given 
by 
$$ 
\dot{\psi}f(\phi) := \left(\frac d{dt}f(\phi+t\psi)\right)_{|t=0}\ ,
$$ 
for any $f\in C^\infty(\cal H)$.


Through the identification of $\cal H$ with $T_\phi\cal H$ at any $\phi\in \mathcal{H}$ a continuous
vector field is just a continuous
map $A\colon \cal H\to \cal H$. In particular,  
a linear operator $A$ on $\cal H$ is a special kind of
vector field. Usually, operators
in Quantum mechanics are neither continuous nor defined on the whole space
$\cal H$.

The most relevant case is when $A$ is a skew-self-adjoint operator, $A=-i\, H$.
The reason is that $\cal H$ can be endowed with
a natural (strongly) symplectic structure, and then such skew-self-adjoint
operators are singularized as the linear vector fields that are
Hamiltonian. The integral curves of such a Hamiltonian vector field  
$A=-i\, H$ are the solutions of the corresponding Schr\"odinger equation
\cite{BCG}. Even when $A$ is not bounded, it can be shown that if $A$ is skew-self-adjoint it must be
densely defined and its integral curves are strongly continuous and defined in all $\mathcal{H}$.


On the other side, the skew-self-adjoint operators considered as vector fields are
fundamental vector fields relative to the usual action of the unitary group
$U(\mathcal{H})$ on the Hilbert space $\mathcal{H}$.

It is important to remark that given a $r$-dimensional real 
Lie algebra of skew-self-adjoint operators $-iH_\alpha$
\begin{equation}
  [iH_\alpha,iH_\beta]=c_{\alpha\beta}\,^{\gamma}\ iH_\gamma,\qquad
  c_{\alpha\beta}\,^{\gamma}\in\mathbb{R}
\label{algebH}
\end{equation}
we can choose a basis $\{a_\alpha\,|\,\alpha=1,\ldots,r\}$ of an abstract Lie
algebra $\goth{g}$ isomorphic to that of the $X_\alpha$ such that the Lie
brackets of the elements $a_\alpha$ of this Lie algebra, denoted by $[\cdot,\cdot]$ satisfy
\begin{equation}
[a_\alpha,a_\beta]=c_{\alpha\beta}\,^\gamma\  a_\gamma\,,,\qquad
  c_{\alpha\beta}\,^{\gamma}\in\mathbb{R}.
\end{equation}

Now, Lie system theory applies to the case in which a $t$-dependent 
Hamiltonian can be written as a linear combination with $t$-dependent 
real coefficients of some self-adjoint operators, 
 \begin{equation}\label{LieHamiltonian}
 H(t)=\sum_{\alpha=1}^rb_\alpha(t)H_\alpha\,,
\end{equation}
where the Hamiltonians $H_\alpha$ are such that the skew-self-adjoint operators
$-iH_\alpha$ close 
a real finite dimensional Lie algebra under the commutator bracket
as indicated in (\ref{algebH}).

When this happens, we can associate a Lie algebra $\LG$ acting on $\cal{H}$ by
 skew-self-adjoint operators in such a way that there exists a basis
 $\{a_\alpha\}$ of $\LG$
with $-iH_\alpha$ being the  fundamental vector field associated with
$a_\alpha$. 

 The linear map  $X:\LG\to  \mathfrak{X}(\mathcal{H})$, $X: a_\alpha\mapsto
 -iH_\alpha$
 is a Lie algebra isomorphism. Then, when $H(t)$ is given by
 (\ref{LieHamiltonian}), 
the Schr\"odinger equation
\begin{equation}
 \frac{d\psi}{dt}=-iH(t)\psi=-\sum_{\alpha=1}^rb_{\alpha}(t)iH_\alpha\psi
\end{equation}
can be reduced to  an equation on $G$ like (\ref{eqingr}) determined by the curve in $\LG$ given
by
$a(t)\equiv{\displaystyle
-\sum_{\alpha=1}^r}b_\alpha(t) a_\alpha$. Once   equation (\ref{eqingr}) with
initial condition $g(0)=e$ has
been  solved we can
obtain the general solution of the Schr\"odinger equation.

As a first instance, we can see how the method works in the very simple
 case of a quantum time-dependent linear potential, which has 
 recently been studied by
Guedes \cite{Gu01}, instead of using the Lewis and Riesenfeld invariant 
method \cite{LR69}. Our method  is  an improvement with respect to previous
ones, 
because it allows us to obtain the solution in an algorithmic way. 

In this case the  quantum Hamiltonian is
\begin{equation}
H_q=\frac{P^2}{2m}+f(t)\, X.\label{Hq}
\end{equation}

In  this quantum problem, as pointed out in  
 \cite{CarRam03, Ba01},  the quantum Hamiltonian
$H_q$ may be written as a sum
$$
H_q=\frac 1 m \, H_1-f(t)\, H_2\ ,
$$
with
$$
H_1=\frac {P^2}2\ ,\qquad  H_2=-X\ .
$$
But  $-i\, H_1$ and $-i\, H_2$  are skew-self-adjoint and close on a four-dimensional Lie algebra 
with $-i\, H_3=-i\,P$, and $-i H_4=i\,I$, which is an extension 
of the opposite of the Heisenberg 
Lie algebra  (\ref{Xla}), i.e. the defining commutation relations are
$$
[-i\,H_1,-i\, H_2]=-i\,H_3\,,\ [-i\,H_1,-i\, H_3]=0\,,\ [-i\,H_2,-i\, H_3]=-i\,H_4\,.
$$

Therefore, the Lie algebra of
 Schr\"odinger equation given by the Hamiltonian $H_q$ is like in the analogous
 classical 
 Lie system. Despite of the fact that $H_q$ 
given by (\ref{Hq}) is time-dependent, it is a Lie system and 
thus we can find the time-evolution operator by solving 
a related equation on the corresponding Lie group 
by the Wei--Norman method.

More explicitly, let $\{a_1,\,a_2,\,a_3,\,a_4\}$ be a basis of the Lie algebra with 
non-vanishing defining relations $[a_1,a_2]=-a_3$ and $[a_2,a_3]=-a_4$.
The  equation  (\ref{eqingr}) in the group to be 
considered is now
$$
R_{g^{-1}*g}\dot g=-\frac 1 m\, a_1+f(t)\, a_2\ .
$$

In order to find the expression of the wave-function in a
simpler way, it is advantageous to use the factorization
$$
g=\exp(-v_4\, a_4)\exp(-v_2\, a_2)\,\exp(-v_3\, a_3)\,\exp(-v_1\, a_1)\,.
$$
In such a case, the Wei--Norman method provides the system
\begin{eqnarray}
&\dot v_1=\dfrac 1m\,, \quad\quad &\dot v_2=-f(t)\,,                     \nonumber\\
&\quad\quad\dot v_3=\dfrac 1m\,v_2\,, \quad\quad &\dot v_4=-\dfrac 1{2m} v_2^2\,,
                                                                \nonumber
\end{eqnarray}
that, jointly with the initial conditions $v_1(0)=v_2(0)=v_3(0)=v_4(0)=0$
determines the solution
\begin{eqnarray}
&&v_1(t)=\frac tm\,,\quad v_2(t)=-\int_0^t dt'\,f(t')\,,\quad           \label{vs12}\\ 
&&v_3(t)=-\frac 1 m \int_0^t dt'\int_0^{t'} dt'' f(t'')\,,               \label{vs3}\\ 
&&v_4(t)=-\frac 1{2m}\int_0^t dt'\left(\int_0^{t'} dt'' f(t'')\right)^2\,. \label{vs4} 
\end{eqnarray}
Then, applying the evolution operator onto the initial wave-function $\psi(p,0)$, which
is assumed to be written in momentum representation, we have
\begin{eqnarray}
\psi_t(p)&=&U(t,0)\psi_0(p)                                                     \nonumber\\
&=&\exp(i v_4(t))\exp(i v_2(t) X)\exp(-i v_3(t) P)\exp(-i v_1(t) P^2/2)\psi_0(p)               \nonumber\\
&=&\exp(i v_4(t))\exp(i v_2(t) X)e^{-i(v_3(t) p+v_1(t) p^2/2)}\psi_0(p)                             \nonumber\\
&=&e^{-i(-v_4(t)+v_3(t) (p-v_2(t))+v_1(t) (p-v_2(t))^2/2)}\psi_0(p-v_2(t))\,,                           \nonumber
\end{eqnarray}
where the functions $v_i(t)$ are given by (\ref{vs12}), (\ref{vs3}) and (\ref{vs4}),
respectively.

We can proceed in a similar way with the quadratic Hamiltonian in the
quantum case, given by \cite{KBW} (see \cite{CarRam03}) 
\begin{equation}
H_q=\alpha(t)\,\frac{P^2}2+\beta(t)\,\frac{X\,P+P\,X}4+\gamma(t)\,\frac{X^2}2+
\delta(t)P+\epsilon (t)\, X+\phi(t)I ,\label{gqH}
\end{equation}
where $X$ and $P$ are the position and momentum operators satisfying the commutation relation 
$$[X,P]=i \, I\ .
$$
It is important to solve this quantum quadratic Hamiltonian because of
its appearance in many branches of physics.

The Hamiltonian can be written as a sum with $t$-dependent coefficients
$$H=\alpha(t)\, H_1+ \beta(t)\, H_2+\gamma(t)\, H_3-\delta(t)\, H_4+\epsilon(t)\, H_5\,+\phi(t)\,H_6
$$
of the Hamiltonians
$$
H_1(x,p)=\frac {P^2}2\,,\quad H_2(x,p)= \frac 14 (XP+P\,X),\quad \
H_3(x,p)=\frac {X^2}2\,,
$$
$$ H_4(x,p)=-P\,,\qquad  H_5(x,p)=X\,, \qquad 
H_6(x,p)=I\,,
$$
which satisfy the commutation relations
\begin{eqnarray}
\begin{aligned}
&[iH_1,iH_2]=iH_1, &[iH_2,iH_3]&=iH_3, &[iH_3,iH_4]&=iH_5, &[iH_4,iH_5]=-iH_6,\\
&[iH_1,iH_3]=2i\, H_2,&[iH_2,iH_4]&=-\frac i2\, H_4,&[iH_3,iH_5]&=0,& \\
&[iH_1,iH_4]=0,&[iH_2,iH_5]&=\frac i2\, H_5\,,\ &&&\\
&[iH_1,iH_5]=-iH_4\,,&& && & \nonumber
\end{aligned}
\end{eqnarray}
and $[iH_\alpha,iH_6]=0$ for $\alpha=1,\dots,5$.

This means that the skew-self-adjoint operators $-i\, H_\alpha$ generate a 
six-dimensional real Lie algebra. Thus, we can define a basis
$\{a_1,\ldots,a_6\}$ with the same structure constants as the $iH_\alpha$ with
Lie product  the commutator of operators.

This six-dimensional real Lie algebra is a central extension 
of the Lie algebra arising in the classical case by the 
one-dimensional Lie subalgebra generated by $a_6$. 
It is a semidirect sum of the Lie subalgebra $\goth{sl}(2,\R)$  
spanned by $\{a_1,a_2,a_3\}$ and the  Heisenberg--Weyl 
Lie algebra generated by $\{a_4,a_5,a_6\}$, which is an ideal.

In full similarity with the classical case, in order  to find the time-evolution provided by the Hamiltonian (\ref{gqH}) 
we should find the curve $g(t)$ in $G$ such that 
$$R_{g^{-1}*g}\dot g=-\sum_{\alpha=1}^6 b_\alpha(t)\, a_\alpha\ ,
\qquad g(0)=e\,,
$$
with 
$$
b_1(t)=\alpha(t)\,,\ b_2(t)=\beta(t)\,,\ b_3(t)=\gamma(t)\,,\ b_4(t)=-\delta(t)\,,\ b_5(t)=\epsilon(t)\,,\ b_6(t)=\phi(t)\,.
$$

This can be carried out by using the generalized Wei--Norman method, i.e. 
by writing $g(t)$ in terms of a set of second class canonical coordinates.  
For instance, 
\begin{eqnarray}\label{factorization}
g(t)&=&\exp(-v_4(t)a_4)\exp(-v_5(t)a_5)\exp(-v_6(t)a_6)\times\cr
&&\times \exp(-v_1(t)a_1)\exp(-v_2(t)a_2)\exp(-v_3(t)a_3)
\end{eqnarray}
and a straightforward application of the above mentioned Wei--Norman
 method technique leads to the system
\begin{eqnarray}\label{WNQH}
\left\{\begin{aligned}
\dot v_1&=b_1+b_2\, v_1+b_3\,v_1^2\ ,&\quad&\dot v_4=b_4+\frac 12\, b_2\, v_4+b_1\,v_5\ ,\cr
\cr
\dot v_2&=b_2+2\,b_3\,v_1\ ,&\quad&\dot v_5=b_5-b_3\, v_4-\frac 12\, b_2\,v_5\ ,\cr
\cr
\dot v_3&=e^{v_2}\,b_3\ , &\quad&\dot v_6=b_6-b_5\, v_4+\frac 12\, b_3\,v_4^2-\frac 12\, b_1\,v_5^2,
\cr
\end{aligned}\right.
\end{eqnarray}
with $v_1(0)=v_2(0)=v_3(0)=v_4(0)=v_5(0)=v_6(0)=0$.

If we consider the following vector fields 
\begin{equation}
\begin{aligned}
X_1&=\pd{}{v_1}+v_5\pd{}{v_4}-\frac{1}{2}v_5^2\pd{}{v_6},\\
X_2&=v_1\pd{}{v_1}+\pd{}{v_2}+\frac{1}{2}v_4\pd{}{v_4}-\frac{1}{2}v_5\pd{}{v_5},\\
X_3&=v_1^2\pd{}{v_1}+2v_1\pd{}{v_2}+e^{v_2}\pd{}{v_3}-v_4\pd{}{v_5}+\frac{1}{2}v_4^2\pd{}{v_6},\\
X_4&=\pd{}{v_4},\\
X_5&=\pd{}{v_5}-v_4\pd{}{v_6},\\
X_6&=\pd{}{v_6}\,,\\
\end{aligned}
\end{equation}
it is a straightforward computation to see that these vector fields close on 
the same commutation relations as the corresponding $\{a_\alpha\}$ and 
thus (\ref{WNQH}) is a Lie system related with the same 
Lie group as the Hamiltonian (\ref{gqH}) or its corresponding 
equation in its Lie group.

Now, once the $v_\alpha(t)$ have been determined, the 
time-evolution of any state will be given by 
\begin{multline}
\ket{\psi(t)}=\exp(-v_4(t)iH_4)\exp(-v_5(t)iH_5)\exp(-v_6(t)iH_6)\times\cr
\times \exp(-v_1(t)iH_1)\exp(-v_2(t)iH_2)\exp(-v_3(t)iH_3)\ket{\psi(0)}\ .\nonumber\\
\end{multline}
and thus
\begin{multline}
\ket{\psi(t)}=\exp(v_4(t)iP)\exp(-v_5(t)iX)\exp(-v_6(t)iI)\times\cr
\times \exp\left(-v_1(t)i\frac{P^2}{2}\right)
\exp\left(-v_2(t)i\frac{PX+XP}{4}\right)
\exp\left(-v_3(t)i\frac{X^2}{2}\right)\ket{\psi(0)}\ ,\nonumber
\end{multline}

\section{The reduction method in Quantum Mechanics}\label{reduction}
We start  this section with a quick review of the reduction technique explained
for example in \cite{CarRamGra,CarRam} and then we obtain some new results. First, as a new improvement, while in some before works like \cite{CarRamGra,CarRam} some sufficient conditions where explained to perform a reduction process, here we show that these conditions can be considered as necessary. Also, we will use the reduction technique to explain the interaction picture used in Quantum Mechanics and we will review from the point of view of our theory the method of unitary transformations.

It has been proved in Section 2 that the study of Lie systems in homogeneous
spaces can be reduced to
that of the solution of equations 
\begin{equation}\label{eqal}
R_{g^{-1}*g} \dot{g}=-\sum_{\alpha=1}^rb_\alpha(t)a_\alpha\equiv a(t)\in T_eG
\end{equation}
with $g(0)=e$. 

The reduction method developed in \cite{CarRamGra} has also shown that given a
solution $\tilde x(t)$ of a Lie system in a homogeneous space $G/H$, 
the solution of the Lie system in the group $G$, and therefore 
the general solution in the given homogeneous space, 
can be reduced to that of a Lie system in the subgroup $H$. 
More specifically,  if the curve $\tilde g(t)$ in $G$ is such that
$\tilde x(t)=\Phi(\tilde g(t),\tilde x(0))$
 with $\Phi$ being the given action of $G$ in the homogeneous space, then
 $g(t)=\tilde g(t)g'(t)$,
 where $g'(t)$ turns out to be a curve in $H$  which 
is a solution of a Lie equation in the
 Lie subgroup $H$ of $G$. 
Actually, once  the curve 
 $\tilde g(t)$ in $G$ has been fixed,  the curve $g'(t)$, which takes values in
 $H$,   
satisfies the equation \cite{CarRamGra}
\begin{equation}\label{trans}
 R_{g'^{-1}*g'}\dot{g'}=-\Ad(\tilde g^{-1})
\left(\sum_{\alpha=1}^rb_\alpha(t)a_\alpha
+R_{\tilde g^{-1}*\tilde g}\dot{\tilde g}\right)
\equiv
a'(t)\in T_eH\,.
\end{equation}

This transformation law can be understood in the language of connections. 
It has been shown in \cite{CarRamGra,CarRam05b} that Lie systems can be related
 with connections in a bundle and that the group of curves in $G$, which is
the group of automorphisms of the principal bundle $G\times\mathbb{R}$ \cite{CarRam05b}, acts on
the left on the set of Lie systems in $G$, and defines an
 induced action on the set of Lie
systems in each homogeneous space for $G$. More specifically, if $x(t)$ 
is a solution of a Lie system in a homogeneous space $M$ defined by the curve
$a(t)$ in $\goth{g}$, then for each curve $\bar g(t)$ in $G$ such that $\bar
g(0)=e$ we see that $x'(t)=\Phi(\bar g(t), x(t))$ is 
a solution of the Lie system defined by the curve 
\begin{equation}
a'(t)=R_{\bar g^{-1}*\bar g}\dot{\bar g} +\Ad(\bar g)a(t)\ .\label{redumet}
\end{equation}
which is the transformation law for a connection.

In summary, the aim of the reduction method is to find an 
automorphism $\bar g(t)$ such that the right-hand side 
in (\ref{redumet}) belongs to $T_eH\equiv\LH$ for a certain Lie
subgroup $H$ of $G$. In this way, 
the papers \cite{CarRamGra,CarRam05b} gave a
sufficient condition
 for obtaining  this result. In this section we shall study the above
 geometrical development
 in Quantum Mechanics and we find out a necessary condition for the right-hand
 side in (\ref{redumet})
 to belong to $\LH$.

Lie systems in Quantum Mechanics are those such that
\begin{equation}
 H(t)=\sum_{\alpha=1}^rb_\alpha(t)H_\alpha\ .
\end{equation}
with  $-iH_\alpha$ closing   under the commutator on a 
finite dimensional real Lie algebra $\goth{v}$. Therefore,
by regarding these operators as fundamental vector fields 
of an action of a connected Lie group $G$ with  
Lie algebra $\LG$ isomorphic to $\goth{v}$, 
we can relate the Schr\"odinger equation with a 
differential equation in $G$ determined by 
curves in $T_eG$ given by 
$a(t)=-{\displaystyle\sum_{\alpha=1}^r}b_\alpha(t)a_\alpha$ 
by considering $-iH_\alpha$ as fundamental vector 
fields of the basis of $\LG$ given by $\{a_\alpha\}$.

Now, the preceding methods allow us to
transform the problem into a new one in  the same group $G$, for each choice
of the curve $\bar g(t)$  but with a new curve $a'(t)$. The action of $G$
on $\mathcal{H}$  is given by an unitary representation $U$, and therefore the time-dependent vector
field determined by the original Hamiltonian $H$ will become a new one with
Hamiltonian $H'$. Its integral curves are the solutions of 
the equation 
\begin{equation}
\frac{d\psi'}{dt}=-iH'(t)\psi'
\end{equation}
where
\begin{equation}
-iH'(t)=-iU(\bar g(t))H(t)U^\dagger(\bar g(t))
+\dot U(\bar g(t))U^\dagger(\bar g(t))
\end{equation}

 That is, from the geometric point of view we 
have related a Lie system on the Lie group $G$ with
 certain curve $a(t)$ in $T_eG$ and the corresponding 
system in $\mathcal{H}$ determined by a unitary 
representation of $G$ with another one with different 
curve $a'(t)$ in $T_eG$ and its associated 
one in $\mathcal{H}$.

Let us choose a basis of $T_eG$
given by $\{c_\alpha\mid \alpha=1,\ldots,r\}$ 
with $r=\dim\, \LG$, such that
$\{c_\alpha\mid \alpha=1,\ldots,s\}$ be a basis of $T_eH$, 
where $s=\dim\,\LH$, 
and denote $\{c^\alpha\mid\alpha=1,\ldots,r\}$ 
the dual basis of $\{c_\alpha\mid
\alpha=1,\ldots,r\}$. In order to find
  $\bar g$  such that the right hand term of (\ref{redumet})
belongs to $T_eH$ for all $t$, the
condition for $\bar g$ is
\begin{equation*}
c^\alpha\left( \Ad (\bar g)a(t)+R_{\bar g^{-1}*\bar g}\dot{\bar g}\right) =0\,,\qquad \alpha=s+1,\ldots,r\,.
\end{equation*}
Now, if $\theta^\alpha$ is the left invariant 1-form on
$G$ induced from $c^\alpha$ the previous equation implies
\begin{equation}\label{condition}
\theta^\alpha_{\bar g^{-1}}
\left(R_{\bar g^{-1}*e}a(t)-\frac{d{\bar g}^{-1}}{dt}\right)=0\,,\qquad \alpha=s+1,\ldots,r\,.\\
 \end{equation}

Let $\tilde g=\bar g^{-1}$, the last expression implies that 
$R_{\tilde g*e}a(t)-\dot{\tilde g}$ is generated by left invariant vector
fields on $G$ from the elements of $\LH$. Then, 
given $\pi^L:G\to G/H$, the kernel of $\pi^L_*$ is spanned 
by the left invariant vector fields on $G$ generated 
by the elements of $\LH$. Then it follows
\begin{equation}
\pi^L_{*\tilde g}(R_{\tilde g*e}a(t)-\dot{\tilde g})=0\,.
\end{equation}

Therefore, if we use that $\pi^L_*\circ X^R_\alpha=-X^L_\alpha\circ\pi^L$,
where $X^L_\alpha$ denotes the fundamental vector field of 
the action of $G$ in $G/H$ and $X^R_\alpha$ denotes 
the right-invariant vector field in $G$ whose value in $e$ is $a_\alpha$, we can prove 
that $\pi^L(\tilde g)$ is a solution on $G/H$ of the equation
\begin{equation}\label{redeq}
\frac{d\pi^L(\tilde g)}{dt}
=\sum_{\alpha=1}^rb_\alpha(t)X_\alpha^L(\pi^L(\tilde g))\,.
\end{equation}

Thus we obtain that given a certain solution $g'(t)$ in $\LH$ related 
to the initial $g(t)$ by means of $\tilde g(t)$ according to 
$g(t)=\tilde g(t)g'(t)$, then the projection to $G/H$ of 
$\tilde g(t)$ is a solution of (\ref{redeq}).

The result so obtained shows that whenever $g'(t)$ 
is a curve in $H$ then $\tilde g(t)$ verifies (\ref{redeq}).
Moreover, as it has been shown in \cite{CarRamGra},
 if $\tilde g(t)$ satisfies 
(\ref{redeq}), then $g'(t)$ is a curve in $H$
satisfying (\ref{trans}). 
The previous result shows that such a condition for 
obtaining (\ref{trans}) is 
not only sufficient but necessary too. Thus, we obtain a new result which 
completes the one found in \cite{CarRamGra}.

Finally, it is worthy of remark that even when this last proof
has been developed for Quantum Mechanics, it can be applied also in ordinary differential equations
because it appears as a consequence of the group structure of Lie systems which
is the same for both the Schr\"odinger and general differential equations.

\subsection{Interaction picture and Lie systems}

As a first new application of the reduction method for Lie systems, we will analyse in this section how this theory can be applied to explain the interaction picture used in Quantum Mechanics. This picture has been shown to be very effective in the developments
of perturbation methods. It plays a r\^ole when the Hamiltonian can be written 
as a sum of a simpler Hamiltonian $H_1$ and another perturbation $H_I$. 
In the framework of Lie systems we can analyze what happens when the Hamiltonian is 
\begin{equation}
H=H_1+V(t)=H_1+\sum_{\alpha=2}^rb_\alpha(t)H_\alpha
=\sum _{\alpha=1}^rb_\alpha(t)H_\alpha,\quad  b_1(t)=1,
\end{equation}
where the set of skew-self-adjoint operators 
$\{-iH_\alpha\, |\, \alpha=1,\ldots,r\}$ is closed 
under commutation and generates a finite dimensional 
real Lie algebra. The situation is very much similar to the case of 
control systems with a drift term (here $H_1$) which are linear in the
controls. The functions $b_\alpha(t)$ correspond to the control functions. 

According to the theory of Lie systems, let us consider 
a basis $\{a_\alpha\, |\, \alpha=1,\ldots,r\}$ of the 
Lie algebra with corresponding associated fundamental 
vector fields $-iH_\alpha$. The equation to be studied 
in $T_eG$ is (\ref{eqal})
and if we define $g'(t)=\bar g(t)g(t)$, where $\bar g(t)$ is a 
previously chosen curve, it holds a similar equation for $g'(t)$
given by (\ref{redumet}).  
%

If in particular we choose $\bar g(t)=\exp(a_1t)$ 
we find the new equation in $T_eG$
\begin{equation}\label{cc}
R_{g'^{-1}*g'}\dot g'=-\Ad(\exp({a_1t}))
\left(\sum_{\alpha=2}^rb_\alpha(t)a_\alpha\right)
=-\exp({\ad(a_1)t})\left(\sum_{\alpha=2}^rb_\alpha(t)a_\alpha\right)\,.
\end{equation}
Correspondingly, the action of $G$ on $\mathcal{H}$ by a unitary 
representation defines a transformation of $\mathcal{H}$ in 
which the state $\psi(t)$ transforms into 
$\psi'(t)=\exp(iH_1t)\psi(t)$ and its dynamical evolution 
is given by the vector field corresponding to the right 
hand side of (\ref{cc}). In particular, 
if $\{a_2,\ldots,a_r\}$ span an ideal of the 
Lie algebra $\LG$ the problem reduces to the 
corresponding normal subgroup in $G$.

\subsection{The method of unitary transformations}

A second application of the theory of Lie systems in Quantum Mechanics and in particular of the reduction 
method is to obtain information about how to proceed to solve a Lie Hamiltonian
in Quantum mechanics. Even when the theoretical results used here can already
be found in the literature, the way to proceed can be considered a new improvement.

Every Schr\"odinger equation of Lie type is determined by a Lie algebra $\LG$,
 a unitary representation of its connected Lie group $G$ in $\mathcal{H}$
and a curve $a(t)$ in $T_eG$. Depending on $\mathfrak{g}$ there are two cases. If $\LG$ is solvable we can use the reduction method in Quantum Mechanics to obtain the general solution. If $\LG$ is not solvable, it is not possible to
integrate the problem in terms of quadratures in the most general case. 
But it may be  possible to solve a problem completely for some specific curves as
for instance 
it happens for the Caldirola--Kanai Hamiltonian \cite{HW98}. A way of dealing
with such systems 
is to try to change the curve $a(t)$ into another one $a'(t)$, 
easier to handle, as it has been done in
the previous subsection of the interaction picture. 
In a more general case than the interaction picture, although any
two curves $a(t)$ and $a'(t)$ are always connected by an automorphism, 
nevertheless, the equation
which determines 
the transformation can be as difficult to be solved as the initial problem. 
Because of this,
it is interesting to
 look for a curve easy enough to be solved but that 
we can connect to the initial
 problem.
 In any case, we can always express the solution of the initial 
problem in terms of a solution of the equation determining the transformation. 
In certain cases, for an appropriate choice of the curve $\bar g(t)$
 the new curve 
$a'(t)$ belongs to $T_eH$ for all $t$, where $H$ is a solvable 
Lie subgroup of $G$. In
this case we can 
reduce the problem from $\LG$ to a certain solvable Lie subalgebra $\LH$ of
$\LG$. Of course, 
in order to do this, a solution of the equation of reduction is
needed, 
but once this is known we can solve the problem completely in terms of it. 
Other methods have alternatively been used in the
literature, like the Lewis-Riesenfeld (LR) method. 
However, this method seems to offer a complete solution 
only if $\LG$ is solvable. If $\LG$ is not solvable, the LR method 
offers a solution which depends on a solution of differential equations, 
like in the method of reduction. 

To sum up, given a Lie system with Lie
algebra $\LG$, with $G$ acting unitarily 
on $\mathcal{H}$ and determined by a curve
$a(t)$ in $T_eG$, 
the systematic procedure to be used is the following: 
\begin{itemize}
\item If $\LG$ is solvable we can solve 
the problem easily by quadratures as it  
appears in \cite{Gu01, Fe01}
\item  if $\LG$ is not solvable, we can try to solve 
the problem for a given curve like in the 
Caldirola--Kanai Hamiltonian in \cite{HW98}, by
choosing
 a curve $\bar g(t)$ transforming
the curve $a(t)$ into another easier to solve, 
like in the interaction picture. 
If this does not work we can try to reduce the
problem, like in the time-dependent mass and 
frequency harmonic oscillator or quadratic
one-dimensional Hamiltonian in \cite{YKUGP94,{CLR08}, So00, FM03} 
to an integrable case.
\end{itemize}

\section{Applications in Quantum Mechanics} 

In this section we shall apply our methods to obtain time-dependent
evolution operators of several problems
found in the physics literature in an algorithmic way. 
In particular, we will analyze several examples of quadratic 
Hamiltonians. These examples are studied in the literature 
under different approaches but here we will study them using 
the same viewpoint. In our description we will classify the 
developed examples in terms of whether the related Lie group
is either solvable or not. 
Also, in the non-solvable cases, we will describe some approaches 
to the study of these Lie systems. 

\subsection{Solvable Hamiltonians}\label{SH}

Quadratic Hamiltonians describe a very large class of physical models. 
Sometimes, one of these physical models is described by a certain 
family of quadratic Hamiltonians that can be considered as a 
Lie system related with a Lie subgroup of the one given for general
quadratic Hamiltonians. When this Lie subgroup is solvable 
the differential equations related with it through the 
Wei--Norman methods are solvable too and the time-evolution 
operator can be explicitly obtained. In this subsection 
we will deal with some instances of this case. In this cases, we can find the 
explicit solution of these problems already the literature by using for each
case 
a different method.

First, we will fix our attention at the motion of a particle 
with a time-dependent mass under the action of a time-dependent 
linear potential term. The Hamiltonian that describe 
this physical case is
\begin{equation}
 H=\frac{P^2}{2m(t)}+S(t)X\,.
\end{equation}

The Lie algebra associated with this example is a central
 extension of the Heisenberg Lie algebra and at the same time,
a Lie subalgebra of the one obtained for quadratic Hamiltonians. 
A basis for the Lie algebra of vector fields related 
with this physical model is
\begin{equation}
Z_1=i\frac{P^2}{2},\quad Z_2=iP,\quad Z_3=iX,\quad, Z_4=iI
\end{equation}
which closes on a Lie algebra under operator commutation,
\begin{equation}
\begin{aligned}
\left[Z_1, Z_2\right]&=0,
\quad&\left[Z_1, Z_3\right]&=2Z_2,
\quad&\left[Z_1,Z_4\right]&=0,\\
\left[Z_2, Z_3\right]&=Z_4,\quad&\left[Z_2, Z_4\right]&=0,\quad&&\\
\left[Z_3, Z_4\right]&=0. &&&&\\
\end{aligned}
\end{equation}
This Lie algebra is solvable, and then, the related equations 
obtained through the Wei--Norman method, 
can be solved by quadratures for any pair 
of time-dependent coefficients $m(t)$ and $S(t)$. 
The solution of the associated Wei-Norman system
allows us to obtain the time-evolution operator 
and the wave function solution of the time-dependent 
Schr\"odinger equation. 

This Hamiltonian has been studied in \cite{UYG02}  for some particular cases 
 using {\it ad-hoc} 
methods
and in general in \cite{Fe01}. Here, we will use 
the Wei--Norman method. As a quadratic Hamiltonian 
its equation in the group $G$ is a particular case 
of the one  related with (\ref{gqH})
\begin{equation}\label{gMS}
R_{g^{-1}*g}\dot g=-\frac 1{m(t)}a_1-S(t)a_5\equiv a_{MS}(t)\,.
\end{equation}
If we use the factorization given in (\ref{factorization})
\begin{multline}
g(t)=\exp(-v_4(t)a_4)\exp(-v_5(t)a_5)\exp(-v_6(t)a_6)\times \\ 
\times \exp(-v_1(t)a_1)\exp(-v_2(t)a_2)\exp(-v_3(t)a_3)
\label{fact}
\end{multline}
 we can solve the Schr\"odinger equation by the Wei--Norman 
method through the set of differential equations (\ref{WNQH}) 
for this particular case
\begin{eqnarray}
&& \dot{v}_1=\dfrac1{m(t)}\,,\quad
\dot{v}_2=0\,,\quad
\dot{v}_3=0 \,,\quad \nonumber \\
&& \dot{v}_4=\dfrac{v_5}{m(t)}\,,\quad
\dot{v}_5=S(t)\,,\quad
\dot{v}_6=-S(t)v_4-\dfrac{v_5^2}{2m(t)}\,,\quad\nonumber
\end{eqnarray}
with initial condition $v_1(0)=v_2(0)=v_3(0)=v_4(0)=v_5(0)=v_6(0)=0$.
The solution of this system can be expressed using 
quadratures because the related group is solvable:
\begin{equation}\label{nonconstmass}
\begin{array}{l}
v_1(t)={\displaystyle\int^t_0}\dfrac{du}{m(u)}\cr
v_2(t)=0\cr
v_3(t)=0\cr 
v_4(t)={\displaystyle\int^t_0}\dfrac{du}{m(u)}
\left( {\displaystyle\int^u_0}S(v)dv\right) \cr
v_5(t)={\displaystyle\int^t_0}S(u)du\cr
v_6(t)=-{\displaystyle\int^t_0}S(u)
\left({\displaystyle \int^u_0}\dfrac{dv}{m(v)}
\left({\displaystyle \int^v_0}S(w)dw\right)\right)du 
-{\displaystyle\int^t_0}\dfrac{du}{2m(u)}
\left({\displaystyle\int^u_0}S(v)dv\right)^2
\end{array}
\end{equation}
and the time-evolution operator is obtained by using 
the last expressions in (\ref{fact}),
\begin{center}
\begin{equation}\label{nonconmasste}
\begin{aligned}
U(g(t))	&=\exp(-v_4(t)iH_4)\exp(-v_5(t)iH_5)\exp(-v_6(t)iH_6)\exp(-v_1(t)iH_1)\\
	&=\exp(v_4(t)iP)\exp(-v_5(t)iX)\exp(-v_6(t)iI)\exp(-iv_1(t){P^2}/{2})
\end{aligned}
\end{equation}
\end{center}

Now, in the case of constant mass, (\ref{nonconstmass}) reads as 
\begin{equation*}
\begin{array}{l}
v_1(t)=\dfrac{t}{m}\cr
v_2(t)=0\cr
v_3(t)=0\cr
v_4(t)=\dfrac{1}{m}{\displaystyle\int^t_0}
\left({\displaystyle \int^u_0}S(v)dv\right)du \cr
v_5(t)={\displaystyle\int^t_0}S(u)du\cr
v_6(t)=-\dfrac{1}{m}{\displaystyle\int^t_0}
\left(S(u){\displaystyle\int^u_0}
\left({\displaystyle\int^v_0}S(w)dw\right)dv\right)du
-\dfrac{1}{2m}{\displaystyle\int^t_0}
\left({\displaystyle\int^u_0}S(v)dv\right)^2du\,,
\end{array}
\end{equation*}
which gives the time-evolution operator 
if we use them in (\ref{nonconmasste})

Now, as we have obtained the time-evolution operator 
for a Hamiltonian related with any curve $a_{MS}(t)$ 
we can consider particular instances of it. 
For example, for the curves with constant mass 
$m$ and $S(t)=q\epsilon_0+q\,\epsilon \,\cos(\omega t)$ 
studied in \cite{Gu01} we obtain
\begin{eqnarray}
&& v_1(t)=\dfrac tm\,,\quad v_2(t)=0\,,\quad v_3(t)=0\,,\quad \nonumber\\
&& v_4(t)=\dfrac{q}{2m\omega^2}
(2\epsilon+\epsilon_0 \omega^2 t^2-2\epsilon \cos(\omega t))\,,\quad
v_5(t)=\dfrac{q}{\omega}(\epsilon_0 \omega t+\epsilon \sin(\omega t))\,,\quad
\nonumber
\end{eqnarray}
and 
\begin{multline}
v_6(t)=\dfrac{-q^2}{12 m\omega^3}
\left(4 \epsilon_0^2\omega^3 t^3
-3\epsilon(\epsilon-4\epsilon_0)\omega t
\right.+\\ 
\left. 3\epsilon(4\epsilon+2 \epsilon_0(\omega^2 t^2-2)
-3\epsilon\cos(\omega t))\sin(\omega t)\right)\,.
\end{multline}

The way to obtain a solution with arbitrary non-constant mass and 
$S(t)=q\epsilon_0+q\epsilon \cos(\omega t)$ was pointed out in
\cite{Gu01} and solved in \cite{Fe01}. From our point of view, 
the most general solution is straightforward 
from (\ref{nonconstmass}) because all cases in 
the literature are particular instances of our  
approach with general functions $m(t)$ and $S(t)$.

Now, we can obtain the wave function solution 
of this system. We know that the wave function 
solution with initial condition $\psi_0$ is
\begin{equation}
\begin{aligned}
\psi_t(x)&=U(g(t))\psi_0(x)\\
&=\exp(iv_6(t))\exp(-v_4(t)iP)\exp(-v_5(t)iX)
\exp\left(-v_1(t)i\frac{P^2}{2}\right)\psi_0(x)
\end{aligned}
\end{equation}
However, expressing the initial wave function $\psi_0$ 
in the momentum space as $\phi_0(p)$ 
the solution is given in the same way as before but with
$U(g(t))$ in the momentum representation.
In this case the solution with initial condition $\phi_0(p)$ is
\begin{equation*}
\begin{aligned}
 \phi_t(p)&=U(g(t))\phi_0(p)\\
&=\exp(-iv_6(t))\exp(v_4(t)iP)\exp(-v_5(t)iX)
\exp\left(-iv_1(t)\frac{P^2}{2}\right)\phi_0(p)\\
&=\exp(-iv_6(t))\exp(v_4(t)iP)\exp(-v_5(t)iX)
\exp\left(-iv_1(t)\frac{p^2}{2}\right)\phi_0(p)\\
&=\exp(-iv_6(t))\exp(v_4(t)iP)
\exp\left(-iv_1(t)\frac{(p+v_5(t))^2}{2}\right)\phi_0(p+v_5(t))\\
&=\exp\left(-iv_6(t)+iv_4(t)p-iv_1(t)
\frac{(p+v_5(t))^2}{2}\right)\phi_0(p+v_5(t))
\end{aligned}
\end{equation*}

\subsection{Non-solvable Hamiltonians and particular instances}

In subsection \ref{SH} the differential equations associated 
to the Quantum Hamiltonians treated there were Lie systems 
related with a solvable Lie algebra. 
Thus, by the Wei--Norman method, 
it has been shown that they were integrable by quadratures. 
When this does not happen it is not easy to obtain a 
general solution for the differential equations. 
Now, we will describe some examples of this type 
of quadratic Hamiltonians. In general we will not 
obtain a general solution in terms of the 
time-dependent functions of the quadratic 
Hamiltonians. Nevertheless, we will show that 
for some instances of them the differential 
equations can be integrated.
Note that explicit solutions of these Hamiltonians cannot generally be obtained
 as our theory explains, but under some integrability conditions
on the coefficients   \cite{CLR08, CL08} the solution can be worked out.

As a first case consider the Hamiltonian 
of a forced harmonic oscillator with 
time-dependent mass and frequency given by
$$
 H=\frac{P^2}{2m(t)}+\frac{1}{2}m(t)\omega^2(t)X^2+f(t)X\,.
$$

This case, either with or without time-dependent frequency, 
has been studied in \cite{YKUGP94, Gu01, Ci1}. 
The equations which describe the solutions of 
this Lie system by the method of Wei--Norman are
$$
\begin{aligned} 
 \dot{v}_1&=\dfrac{1}{m(t)}+m(t)\omega^2(t)v_1^2\cr
\dot{v}_2&=2m(t)\omega^2(t)v_1\cr
\dot{v}_3&=e^{v_2}m(t)\omega^2(t)\cr
\dot{v}_4&=\dfrac{1}{m(t)}v_5\cr
\dot{v}_5&=f(t)-m(t)\omega^2(t)v_4\cr
\dot{v}_6&=\dfrac{1}{2}m(t)\omega^2(t)v_4^2-f(t)v_4-\dfrac{1}{2m(t)}v_5^2\nonumber
\end{aligned}
$$
with initial conditions $v_1(0)=v_2(0)=v_3(0)=v_4(0)=v_5(0)=v_6(0)=0$,
where we have used the factorization (\ref{factorization}). 
The solution of this system cannot be obtained by quadratures 
in the general case because the associated 
Lie group is not solvable. Nevertheless, we can consider 
a particular instance of this kind of Hamiltonian, 
the so called Caldirola--Kanai Hamiltonian \cite{HW98}. 
In this case, for the particular time-dependence 
$m(t)=e^{-rt}m_0$, $\omega(t)=\omega_0$ and $f(t)=0$,
the Hamiltonian is given by
\begin{equation}
 H=\frac{P^2}{2m_0}e^{rt}+\frac{1}{2}m_0e^{-rt}\omega_0^2X^2\,.
\end{equation}

In this case the solution is completely known and is given by
\begin{eqnarray}
v_1(t)&=&\dfrac{2e^{rt}}{m_0(r+\bar \omega_0\,{\rm coth}\,(\frac{t}{2}\bar \omega_0))}
\,, \nonumber\\
v_2(t)&=&r t+2\log{\bar \omega_0}-2\log\left(r\sinh\left(\frac{t}{2}\bar \omega_0\right)
+\bar w \cosh\left(\frac{t}{2}\bar \omega_0\right)\right)\,,\nonumber\\
v_3(t)&=&\dfrac{2 m_0 \omega_0^2}{r+\bar \omega_0\,{\rm coth}\,(\frac{t}{2}\bar \omega_0)}\,,
\nonumber\\
v_4(t)&=&0\,,\quad v_5(t)=0\,,\quad v_6(t)=0\,,\nonumber
\end{eqnarray}
where $\bar{\omega}_0=\sqrt{r^2-4 \omega_0^2}$. 
This example shows that the problem may also be 
solved exactly for particular instances of curves 
in $\LG$ of Lie systems with non solvable Lie algebras. 
Another example is the following one
\begin{equation}
 H=\frac{P^2}{2m}+\frac{m \omega_0^2}{2(t+k)^2}X^2\,,
\end{equation}
for which the solution of the Wei--Norman system reads
\begin{eqnarray}
v_1(t)&=&\dfrac{2(k+t)((k+t)^{\bar \omega_0}-k^{\bar \omega_0})}
{m(k^{\bar \omega_0}(\bar \omega_0-1)+(k+t)^{\bar \omega_0}(\bar \omega_0+1))}\,,
\nonumber\\
v_2(t)&=&(1+\bar \omega_0)\log(k+t)-(1+\bar \omega_0)\log k
+2\log(2 k^{\bar \omega_0}\bar \omega_0)\nonumber\\
& &-2\log(k^{\bar \omega_0}(\bar \omega_0-1)+(k+t)^{\bar \omega_0}(\bar \omega_0+1))\,, \nonumber\\
v_3(t)&=&\dfrac{2 m \omega_0^2}{k}\dfrac{(k+t)^{\bar \omega_0}-k^{\bar \omega_0}}
{k^{\bar \omega_0}(\bar \omega_0-1)+(k+t)^{\bar \omega_0}(\bar \omega_0+1)}\nonumber\\
v_4(t)&=&0\,,\quad v_5(t)=0\,,\quad v_6(t)=0\,,\nonumber
\end{eqnarray}
where now $\bar \omega_0=\sqrt{1-4 \omega_0^2}$.

Other examples of 
Hamiltonians, which can be studied by our method, can
be found in \cite{HW98}. We just mention two examples 
which can be completely solved
\begin{eqnarray*}
 H_1&=&\frac{P^2}{2m_0}+\frac{1}{2}m_0(U+V\cos(\omega_0 t))X^2\,,\\ 
H_2&=&\frac{P^2}{2m_0}e^{rt}+\frac{1}{2}m_0e^{-rt}\omega_0^2X^2+f(t)X\,.
\end{eqnarray*}
The first one corresponds to a Paul trap as has been 
studied in \cite{FW95}, and admits a solution in terms of 
Mathieu's functions. The second one is a damped Caldirola--Kanai
Hamiltonian analysed in \cite{UYG02}.

\subsection{Reduction in Quantum Mechanics}

Quite often, when a non-solvable Lie algebra is involved 
in a quantum problem, it is
interesting to solve it in terms of (unknown) solutions of differential
equations. Next, we study some examples of how to
proceed with the method of reduction in order to deal with 
problems in this way. So, we will obtain that the reduction method not only can
be applied in Quantum Mechanics but also allows us to solve certain problems in an algorithmic way. As far as we know, this kind of application of the theory of reduction in Lie systems to a Sch\"odinger equation is new.

Consider an harmonic oscillator with time-dependent 
frequency whose Hamiltonian is given by
\begin{equation}\label{hodt}
 H=\frac{P^2}{2}+\frac{1}{2}\Omega^2(t)X^2\,.
\end{equation}
As a particular case of the Hamiltonian described 
in  Section \ref{SLSQM} this example is related 
with an equation in the connected Lie group associated 
to the semidirect sum of $\goth{sl}(2,\R)$ with the 
Heisenberg Lie algebra generated by the ideal 
$\{a_4,a_5,a_6\}$
\begin{equation}\label{inieqG}
R_{g^{-1}*g}\dot g=-a_1-\Omega^2(t)a_3,\quad g(0)=e.
\end{equation}
Since the solution of this equation starts from the identity
and $\{a_1, a_2, a_3\}$ close on a $\goth{sl}(2,\R)$ Lie algebra,
then the Hamiltonian $H$ of (\ref{hodt}) is related 
with the group $SL(2,\R)$.  Actually this is due to the isomorphism of such a
group with the symplectic one, and in higher dimension the group will be
$Sp(2n,{\mathbb{R}})$
  instead of $SL(2n,\mathbb{R})$.

As a particular application of the reduction technique 
we will perform the reduction from $G=SL(2,\R)$ to the Lie 
group related with the Lie subalgebra 
$\LH=\langle a_1\rangle$. To obtain such a reduction, 
we have shown in Section~\ref{reduction} that we 
have to solve an equation in $G/H$, namely
\begin{equation}
\frac{d\pi^L(\tilde g)}{dt}
=\sum_{\alpha=1}^3b_\alpha(t)X_\alpha^L(\pi^L(\tilde g))
\end{equation}
where $X^L_\alpha$ are the fundamental vector fields 
of the action $\lambda$ of $G$ on $G/H$. 
Now, we are going to describe this equation in a 
set of local coordinates. 
First, in an open neighborhood $U$ of $e\in G$ we can 
write in a unique way any element of $SL(2,\R)$ as
\begin{equation}\label{des}
g=\exp(\alpha_3a_3)\exp(\alpha_2a_2)\exp(\alpha_1a_1).
\end{equation}
where we choose
\begin{equation}
\begin{aligned}
a_1=\left(\begin{array}{cc}
0&1\\
0&0\end{array}
\right),\quad
a_2=\frac 12\left(\begin{array}{cc}
1&0\\
0&-1\end{array}
\right),\quad
a_3=\left(\begin{array}{cc}
0&0\\
-1&0\end{array}\right).&
\end{aligned}
\end{equation}

This decomposition allows us to establish a local 
diffeomorphism between an open neighborhood $V\subset G/H$ 
and the set of matrices given by $\exp(\alpha_3a_3)\exp(\alpha_2a_2)$. 
Now, the decomposition (\ref{des}) reads in matrix terms as
\begin{equation}
\left(
\begin{array}{cc}
\alpha &\beta\\
\gamma &\delta
\end{array}
\right)=
\left(
\begin{array}{cc}
1 &0\\
\phi &1
\end{array}
\right)
\left(
\begin{array}{cc}
\theta &0\\
0 &\theta^{-1}
\end{array}
\right)
\left(
\begin{array}{cc}
1 &\psi\\
0 &1
\end{array}
\right)=
\left(
\begin{array}{cc}
\theta &0\\
\phi\theta &\theta^{-1}
\end{array}
\right)
\left(
\begin{array}{cc}
1 &\psi\\
0 &1
\end{array}
\right)\,.
\end{equation}
If we express $\phi ,\theta, \psi$ 
in terms of $\alpha, \beta, \gamma$ and $\delta$ we obtain
\begin{equation}
\left(
\begin{array}{cc}
\alpha &\beta\\
\gamma &\delta
\end{array}
\right)=
\left(
\begin{array}{cc}
1 &0\\
\gamma/\alpha &1
\end{array}
\right)
\left(
\begin{array}{cc}
\alpha &0\\
0 &\alpha^{-1}
\end{array}
\right)
\left(
\begin{array}{cc}
1 &\beta/\alpha\\
0 &1
\end{array}
\right)=
\left(
\begin{array}{cc}
\alpha &0\\
\gamma &\alpha^{-1}
\end{array}
\right)
\left(
\begin{array}{cc}
1 &\beta/\alpha\\
0 &1
\end{array}
\right)\,.
\end{equation}
Thus, we can consider the projection 
$\pi^L:U\subset G\rightarrow G/H$ given by
\begin{equation}
\pi^L\left(
\begin{array}{cc}
\alpha &\beta\\
\gamma &\delta
\end{array}
\right)=
\left(
\begin{array}{cc}
\alpha &0\\
\gamma &\alpha^{-1}
\end{array}
\right)H
\end{equation}
which allows to represent locally the elements of $G/H$ 
as the $2\times 2$ lower triangular matrices with 
determinant equal to one\,.
Now, given $\lambda_g:g'H\in G/H\rightarrow gg'H\in G/H$ 
as $\lambda_g\circ\pi^L=\pi^L\circ L_g$ then the 
fundamental vector fields defined in $G/H$ 
by $a_1$ and $a_3$ through the action of $G$ on $G/H$ are given by
\begin{equation}
\begin{aligned}
X_1^L(\pi^L(g))&=\frac{d}{dt}\bigg|_{t=0} \pi^L\left(\exp(-ta_1)\left(
\begin{array}{cc}
\alpha &\beta\\
\gamma &\delta
\end{array}
\right)\right)=\left(
\begin{array}{cc}
-\gamma &0\\
0&\gamma/\alpha^2
\end{array}
\right)\\
X_3^L(\pi^L(g))&=\frac{d}{dt}\bigg|_{t=0} \pi^L\left(\exp(-ta_3)\left(
\begin{array}{cc}
\alpha &\beta\\
\gamma &\delta
\end{array}
\right)\right)=\left(
\begin{array}{cc}
0&0\\
\alpha &0
\end{array}
\right)
\end{aligned}
\end{equation}
and the equation in $V\subset G/H$ is described by
\begin{equation}
\begin{aligned}
\left(
\begin{array}{cc}
\dot\alpha &0\\
\dot\gamma &-\dot\alpha\alpha^{-2}
\end{array}
\right)=
\left(
\begin{array}{cc}
-\gamma &0\\
\Omega^2(t)\alpha&\gamma\alpha^{-2}
\end{array}
\right)
\end{aligned}
\end{equation}
and thus we want to obtain a solution of the system
\begin{eqnarray}\label{Oscill}
\left\{\begin{aligned}
\ddot \alpha&=-\Omega^2(t)\alpha\\
\gamma&=-\dot\alpha\,.
\end{aligned}\right.
\end{eqnarray}
Then, if $\alpha_1$ is a solution of the system (\ref{Oscill}) 
the curve $\tilde g(t)$ that verifies $g(t)=\tilde g(t)h(t)$, 
where $h(t)$ is a solution of an equation defined 
in the group related with $\LH=\langle a_1\rangle$, reads
\begin{equation}
\tilde g(t)=
\left(
\begin{array}{cc}
\alpha_1 &0\\
-\dot\alpha_1&\alpha_1^{-1}
\end{array}
\right)
\end{equation}
and the curve which acts on the initial equation 
in $SL(2,\R)$ to transform it
into one in the mentioned Lie subalgebra
is given by $\bar g(t)=\tilde g^{-1}(t)$,
\begin{equation}
\bar g(t)=\exp(-2\log \alpha_1\,a_2)
\exp\left(-\frac{\dot \alpha_1}{\alpha_1}\,a_3\right)
=\exp(-\alpha_1\dot\alpha_1 a_3)\exp(-2\log \alpha_1\,a_2)
\end{equation}
and this curve transforms the initial 
equation in the group given by (\ref{inieqG}) 
into the new one given by (cf. (\ref{redumet}))
\begin{equation}
{a}'(t)=-\frac{1}{\alpha_1^2(t)}a_1\,,
\end{equation}
which corresponds to the Hamiltonian
\begin{equation}
H'=\frac{1}{2\alpha_1^2(t)}P^2\,,
\end{equation}
and the induced transformation in the Hilbert 
space $\mathcal{H}$ that transforms the initial 
$H$ into $H'$ is
\begin{equation}
\exp\left(i\frac{\log \alpha_1}{2}(PX+XP)\right)\exp\left(-i\frac{\dot\alpha_1}{2\alpha_1}X^2\right).
\end{equation}
Both results  can be found in \cite{FM03}. 

There are other possibilities of choosing different  
Lie subalgebras of $\LG$ in order to perform the reduction, 
however the results are always given in terms of a solution
of a differential equation.

\section{Conclusions and outlook}

It has been shown that the geometric approach to Lie systems of differential
equations can be extended to the framework of quantum mechanics and several
examples of application have been developed. In particular
we have studied time-dependent quadratic Hamiltonians for a single particle and
other related models. Some of them are involved in models for
dissipative Quantum Mechanics or harmonic oscillators in external
fields. In all of these applications we have 
developed methods to obtain exact solutions
that summarize many of the different techniques used in the literature in a
 simple and unifying framework. The method of reduction has also been revisited 
and  a new theoretical result and some new applications in Quantum Mechanics
have been obtained.

But these are not the only applications of the methods developed here, 
they can be used in many other applications. For example, 
Hamiltonians of the following type
are studied in Quantum Optics \cite{Yu76}
\begin{multline*}
H=\w_1\hat{a}^\dagger_1\hat{a}_1+\w_2\hat{a}^\dagger_2\hat{a}_2
+\Omega_0(t)(e^{i\w t}\hat{a}^\dagger_1\hat{a}_2
+e^{-i\w t}\hat{a}^\dagger_2\hat{a}_1)\\
+\Omega_L(t)\hat{a}^\dagger_1+\Omega^*_L(t)\hat{a}_1
+\Omega_R(t)\hat{a}^\dagger_2+\Omega_R^*(t)\hat{a}_2\,.
\end{multline*}
In this case, it can be shown that this Hamiltonian is a Lie system and thus we
may study systems of interacting harmonic oscillators in external fields in our
formalism. Many particular cases of this Hamiltonian have
been  studied in several  papers and in different contexts. 

Another different example is a system of coupled spins 
in magnetic fields like the following one found in \cite{YYL99}
\begin{equation*}
 H=-2A\sum_{i,j}{\bf \hat{S}_i}\cdot{\bf \hat{S}_j}
+{\bf B}(t)\cdot\sum_i{\bf \hat{S}_i}
\end{equation*}

In this example our method allows us to obtain exact 
solutions when ${\bf B}(t)$ is a three dimensional 
vector field in the form 
$$
{\bf B}(t)=(B\sin\theta\cos(\omega t), 
B\sin\theta\sin(\omega t), B\cos(\theta))\,.
$$ 

On the other hand, our viewpoint allows us to study important properties of
these and other physical systems too. For example, the last example 
is related with Berry's phases. Our
method allows us to obtain the time-evolution operator. 
Then this information can be used to determine
Berry's phases.  More specifically, the integrability conditions obtained in 
\cite{CLR08} can be generalised to solve exactly some Hamiltonians for certain time-dependent coefficients.
We can also try to apply control theory 
to determine the relation of Berry's phases with the 
field ${\bf B}(t)$, study the adiabatic approximation, etc. 
All these topics are important and will be analyzed 
in forthcoming papers.

\section*{Acknowledgements}

Partial financial support by research projects MTM2006-10531 and E24/1 (DGA)
 are acknowledged. JdL also acknowledges
 a F.P.U. grant from  Ministerio de Educaci\'on y Ciencia.

\end{document}